# Optical Frequency Combs Generated by Four-Wave Mixing in Optical Fibers for Astrophysical Spectrometer Calibration and Metrology

Flavio C. Cruz

*Gleb Wataghin Physics Institute, University of Campinas - UNICAMP*
*13083-970, Campinas, SP, Brazil*
*flavio@ifi. unicamp.br*

**Abstract:** Optical frequency combs generated by multiple four-wave mixing of two stabilized single-frequency lasers in optical fibers are proposed for use as high precision frequency markers, calibration of astrophysical spectrometers and metrology. Use of highly nonlinear and photonic crystal fibers with very short lengths and small group velocity dispersion, combined with energy and momentum conservation required by the parametric generation, assures negligible phase mismatch between comb frequencies. In contrast to combs from mode-locked lasers or microcavities, the absence of a resonator allows large tuning of the frequency spacing from tens of gigahertz to beyond teraHertz.

## 1. Introduction

Direct measurement of optical frequencies was a long standing problem in science and technology, which was elegantly solved in 1999 with the advent of optical frequency combs (OFC) based on femtosecond mode-locked lasers [1], [2]. Since then, these combs allowed for example the development of optical atomic clocks [3],[4], tests of possible variations of fundamental constants [5], extreme nonlinear optics [6], sensitive, high resolution and broadband spectroscopy [7],[8] [9], and high harmonic generation of attosecond pulses [10]. The development of OFC was based on the availability of two technological ingredients: femtosecond lasers in the near infrared, and the development of photonic crystal fibers [11]. These fibers allowed a phase-coherent spectral expansion of the comb (supercontinuum generation), required for optical frequency measurements. Later, ultrabroadband femtosecond lasers [12] were developed that dispensed the use of these fibers, and improved long-term operation. Recently, OFC at 1.5 μm with frequency spacing higher than 88 GHz were generated by four wave mixing (FWM) and Kerr nonlinearity in monolithic high-Q silicon microcavities [13], [14]. These combs require a single input low-power cw laser, and generate new frequencies by four-wave mixing, over demonstrated bandwidths of 500 nm. OFC with line spacing of a few tens of gigahertz have also been demonstrated by spectral filtering of GHz mode-locked lasers, producing frequency spacing between 10 and 40 GHz. Such combs have been proposed for high precision calibration of astrophysical spectrographs in the near infrared [15], [16]. High-precision and long-term stability of such "astro-combs" can increase the accuracy in velocity determination, allowing the search for terrestrial mass planets orbiting distant stars. All of the above implementations of OFC require a resonator, whose physical size fixes the line spacing and allows very limited tuning.

In this paper**,** it is proposed that OFC generated by multiple four wave mixing of two stabilized single-frequency lasers in very short, low-dispersion and highly nonlinear optical fibers can overcome this limitation, providing a stable and precise comb of optical frequencies with tunable spacing, which can be well suited for astrophysical spectrometer calibration, direct frequency comb spectroscopy, and frequency metrology. It is discussed a particular implementation that involves an optical frequency standard with known frequency, as input laser, and a second input laser phase-locked to it using a stable microwave oscillator. Frequency spacings of a few tens of GHz, suitable for astrophysical spectrometer calibration, are readily attainable. The combination of very short lengths, high nonlinearity, and low dispersion over the entire comb bandwidth, assures negligible phase mismatch between the comb frequencies and generation of near transform-limited pulses. This last feature also allows broadband frequency doubling [17] into the near-infrared, which might be required by astrophysical applications [18].

## 2. Accuracy and stability of four-wave mixing fiber combs

The generation of a stable and precise comb of optical frequencies by FWM in an optical fiber requires at least three conditions: 1) the frequency and phase stability of the input lasers must be preserved at the output of the fiber. Their linewidths must not be degraded by technical noise added by the fiber; 2) the nonlinear process responsible for generating new frequencies in the fiber must be coherent, assuring that the new frequencies are phase coherent with the input lasers; 3) dispersion in the fiber should be negligible (or at least known and controllable), so that it will not affect the coherence among the comb lines. Condition 1 can be satisfied by using a short length of optical fiber, well isolated from environmental noise, or by employing well established active techniques of fiber noise cancellation [19,20,21]. In this case a low power, frequency shifted portion of one input laser, sent through the fiber, is reflected back and then heterodyned with the input beam in order to produce an error signal for cancellation of fiber noise. This scheme can be applied for fiber lengths up to a several tens kilometers, since for longer fibers higher propagation delays reduce the servo bandwidth. Condition 2 is intrinsically met by FWM, as consequence of energy and momentum conservation required by the nonlinear generation [22], [13]. Condition 3 can be satisfied with use of short lengths of low-dispersion fibers. A. Cerqueira *et al* reported broadband generation of cascaded FWM products spanning over 300 nm by using very short highly nonlinear optical fibers (HNLF) [23]. As opposed to combs based on mode-locked lasers or microresonators, the absence of a cavity in FWM-combs implies that dispersion adds a phase shift in the comb lines and not a frequency shift in the spectrum. The phase mismatch is given by $\Delta\phi = \Delta\beta.L$, in which $\Delta\beta = \beta_2\Delta\omega^2$ is the dispersion mismatch, where $\beta_2$ is the group velocity dispersion and $\Delta\omega = 2\pi\Delta f$ is the frequency difference among adjacent lines [22]. The phase mismatch between adjacent comb lines (separated by 750 GHz) at $\lambda \approx 1550$ nm, for a 2-m long dispersion-flattened fiber with dispersion parameter $D = -(\lambda/c)d^2n/d\lambda^2 = -(2\pi c/\lambda^2)\beta_2 \approx 1$ ps/nm.km, is $\Delta\phi \approx 40$ mrad. For a comb with 300 nm bandwidth, the phase shift between extreme lines will be $\Delta\phi \approx 100$ rad, and thus still $<\pi$. Fibers with lower dispersion can be employed. For very short fibers, Brillouin scattering is also negligible.

FWM has been extensively studied, particularly in optical fibers [22] since it can be a basic limitation to the operation of fiber-based optical telecommunications systems. In a special case known as multiple FWM, two incident lasers with frequencies $f_1$ and $f_2$ (with $f_2>f_1$), generate a comb of frequencies separated by $\Delta f = f_2-f_1$. Pulses produced by compression of the original beat note are obtained. Several groups have been investigating this technique with a primary interest on generation of short, high-repetition rate pulses for optical time division multiplexing (OTDM) for optical telecommunications [24]. FWM-combs have been demonstrated using optical fibers with anomalous [25] and normal dispersion [26]. Pulses well separated with no pedestal, at rates extending from 20 GHz to up to 1 THz, with durations from 280 fs to a few ps have been demonstrated [26, 27, 28,29].

Recently FWM-combs spanning 300 nm, with 740 GHz spacing, were generated by injecting two lasers into highly nonlinear fibers with lengths of only 2 m [23] (it is interesting to note that the distance traveled by light inside high-Q ($10^8$) microresonators [13] for generation of "Kerr combs" is 24 m). In ref. [23], two independent single-frequency cw lasers at 1555 and 1561 nm were injected into the HNLF after amplification into fiber amplifiers (EDFAs) to peak powers of 10 W. To avoid saturation of the EDFAs, both lasers were switched into 40 ns pulses with low duty cycle. The power in the comb lines varies from -30 dBm (1 μWatt) for frequencies in the wings, to -10 dBm (100 μW) for central frequencies.

Although FWM-combs resemble those generated by mode-locked lasers, pulse trains generated by FWM can have special features such as near transform-limited pulses (e.g. with near zero spectral phase, a consequence of the low dispersion in the fiber) and fixed pulse-to-pulse temporal phase. In mode-locked lasers, a short pulse is formed, for example in a Ti:sapphire laser crystal, by a combination of self-phase modulation (which creates new frequencies) and Kerr lens mode-locking (which locks the phase of these frequencies and favors pulse over cw oscillation). As this pulse propagates inside the cavity, dispersion shifts the carrier frequency with respect to the pulse envelope. After a round trip time T, a phase $\delta\phi$ is acquired between the carrier and the envelope. Because this pulse circulates in the cavity (being damped by the output coupler and amplified by the gain medium), this phase-shift $\delta\phi$ is acquired between successive output pulses, giving rise to a frequency shift $f_{ceo} = \delta\phi/2\pi T$ in the comb spectrum with respect to the cavity resonances. In FWM-combs, however, dispersion in the fiber gives rise to a phase shift which is the same for each output pulse, but not a phase shift between successive pulses. The absence of a cavity does not lead to any frequency shift. There is however a phase shift between successive pulses, which comes from the original beatnote and is equal to π [27, 28]. The beating between the two injected frequencies $f_1$ and $f_2$ consists of their sum frequency modulated by the difference frequency, which causes a π phase difference from one beatnote to the next. New frequencies, generated in the fiber as linear combinations of the two input frequencies, have the effect of narrowing the beatnote, while still keeping the π phase difference [25-28]. The analogy with frequency combs generated by mode-locked lasers can be explored further. In a mode-locked laser, the $n^{th}$ tooth of the comb has a frequency $f_n$ given by:

$$f_n = n\, f_{rep} + f_{ceo} \qquad (1)$$

where $f_{rep}$ is the repetition rate of the laser, n is an integer and $f_{ceo} = \delta\phi/2\pi T = \delta\phi f_{rep}/2\pi$. Both $f_{rep}$ (which is set by the laser cavity length) and $f_{ceo}$ are typically RF or microwave frequencies, ranging from tens of MHz to about 1 GHz. For optical frequency measurements, both $f_{rep}$ and $f_{ceo}$ are actively stabilized to a RF or microwave frequency standard, so that $f_n$ can be known once n is determined by a wavelength measurement, using for example a wavemeter. Furthermore, an optical clock is realized if $f_n$ is locked to an optical frequency standard. By setting $f_{ceo}=0$, $f_{rep}$ will be the RF or microwave output frequency of such clock, having stability and accuracy superior to microwave clocks. In FWM-combs, the $m^{th}$ comb line can be written as:

$$f_m = m\, \Delta f + f_{Offset} \qquad (2)$$

where m is integer. Equation 2 is formally identical to eq. (1), but its completely different meaning simply expresses the fact that any frequency can be written as a multiple of a lower frequency plus an offset. This offset frequency $f_{Offset}$ is simply another microwave (or THz, depending on Δf) frequency, by definition smaller than Δf. It turns out that $f_{Offset}$ can be measured with the same techniques used to measure $f_{ceo}$ (Eq. 1) in mode-locked lasers [30]. A convenient way is when the comb spectrum covers one octave. By knowing $f_{Offset}$, and the difference frequency Δf, given by the microwave reference oscillator, the frequency of laser 1 (if it is unknown) can be determined with respect to the microwave oscillator, using eq. (2).

## 3. Implementation of "astro-combs" of precise optical frequencies

The experimental implementation of a FWM-comb depicted in Fig. 1 can be used to generate a comb of precise optical frequencies, particularly with the required spacing for astrophysical applications. A stable laser with known frequency $f_1$ (optical frequency standard, such as a diode laser stabilized to acetylene or its second harmonic stabilized to Rb) is one of the input lasers, while the other laser (frequency $f_2$) is phase-locked to it using a reference microwave oscillator (standard). It is assumed that accuracy and stability are higher for the optical standard than for the microwave standard. In this case, $f_1$ will be a stable frequency at the fiber output, while all other comb frequencies will have fluctuations approximately given by those of the microwave oscillator. For astrophysical spectrometer calibration, fractional stability and accuracy near $10^{-11}$ would be sufficient to measure velocity variations of 1 cm/s [15].

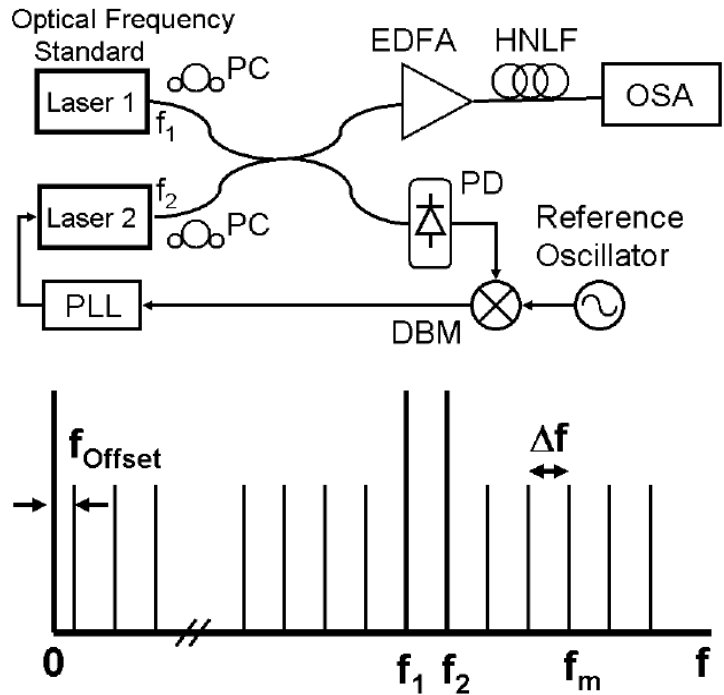

**Fig 1.** Schematic diagram of the experimental setup to generate an OFC by FWM in a highly nonlinear fiber (HNLF). Laser 1 (frequency $f_1$), is an optical frequency standard and laser 2 (at $f_2$) is phase-locked to it using a stable reference oscillator. PC: polarization controller, EDFA: Erbium doped fiber amplifier, PD: photodetector; DBM: double-balanced mixer, PLL: phase-locked loop, OSA: optical spectrum analyzer.

The frequency fluctuations of the two independent free-running input lasers would, in principle, propagate to the new frequencies as they are generated as linear combination of the input frequencies $f_1$ and $f_2$. The fluctuations of the $m^{th}$ comb line would be $\approx m[(\delta f_1)^2+(\delta f_2)^2]^{1/2}$, where $\delta f_1$ and $\delta f_2$ are the fluctuations of the input lasers. This, however, happens only if the length of the fiber is bigger than the coherence length of the input lasers or, in other words, if the transit time through the fiber is longer than the time scale of the laser frequency excursions. Otherwise the fiber will work like a high-pass filter with a cutoff frequency given by the inverse of the transit time (= 100 MHz for a 2-m fiber). Even for free-running telecom lasers with typical linewidths of 100 kHz, the fastest frequency fluctuations occur in a time t ≈ 1/$\delta$f = 1/100kHz = 10 μs, which is much longer than the transit time through the fiber, $t_t$ = 2 m/(c/n) = 10 ns, with n=1.5. As a consequence, the frequency fluctuations of the input lasers (from technical noise, dominated by low-frequency contributions and seldom with components over a few MHz) will not propagate to the new comb frequencies due to the FWM process, as long as parametric generation takes place in a short fiber length. Stabilized lasers or optical standards with very small linewidths, would allow use of much longer fibers. The setup of Fig. 1 should therefore produce a comb of known optical frequencies separated by the adjustable difference $\Delta$f, and with fractional instabilities that should be smaller than $10^{-11}$, and set by the microwave reference oscillator noise. Another obvious implementation of a FWM-comb involves two optical frequency standards, with the frequency spacing instability set by their combined instabilities. In this case, even assuming full noise propagation into the generated comb lines, fractional instabilities <$10^{-11}$ would be attainable.

It is worth noting that near zero spectral phase can be obtained for combs generated by FWM in fibers, allowing optical pulses which are nearly Fourier transform-limited [26]. This should allow the generation of broadband second harmonic spectra [17], thus extending FWM combs into the near-infrared. Other applications of FWM-combs may include direct link between optical and TeraHertz frequencies, secure communications, optical frequency synthesis, and sensitive and broadband spectroscopy [7], [8], [9], with possibility of tunable frequency spacing and of exploring the coherent accumulation effect [31] with alternated pulses at high repetition rates, for driving transitions with short-lived states. A recent proposal for selective excitation of cold molecules also demands pulses at 2 THz rates [32].

## 4. Conclusion

Optical frequency combs based on multiple four wave mixing in short, low-dispersion, highly nonlinear fibers are proposed for use as precise frequency markers and metrology. A distinct feature of such combs is the tunable frequency spacing, which can be set from tens of GHz to beyond one THz. In particular, spacings of a few tens of GHz are suited for calibration of astrophysical spectrometers. A stable comb of precise and known optical frequencies can be generated using an optical frequency standard and a second laser phase-locked to it. Fiber technical noise can be actively cancelled (or minimized with use of short fiber lengths, isolated from environmental noise), preserving the stability of the input lasers. The phase coherent process of FWM should preserve the precision and stability of the input lasers into the new generated frequencies. The expected performance of FWM fiber combs should still be demonstrated, with additional work being surely stimulated by their potential for compactness, reduced cost, long-term operation, and integration with telecom technology.

**Acknowledgments**: The author thanks Arismar Cerqueira Sodre Junior, Hugo L. Fragnito and Christiano de Matos for discussions on multiple FWM and highly nonlinear optical fibers. Financial support from FAPESP, CEPOF, CNPq is gratefully acknowledged.